\documentstyle[preprint,aps,epsf,floats]{revtex}    % PREPRINT STYLE
\input{psfig}
\def\bq{\begin{equation}}
\def\eq{\end{equation}}
\def\ba{\begin{eqnarray}}
\def\ea{\end{eqnarray}}

\begin{document}
\thispagestyle{empty}
\tightenlines

\newcommand{\pslasht}{p\llap/_T}

\preprint{
\font\fortssbx=cmssbx10 scaled \magstep2
\hbox to \hsize{
\hskip.5in \raise.1in\hbox{\fortssbx University of Wisconsin - Madison} 
\hfill\vtop{\hbox{\bf MADPH-97-1017}
            \hbox{hep-ph/9709506} 
            \hbox{September 1997}} }
}
  
\title{\vspace{.5in}
Searching for a heavy Higgs boson via the $H\to l\nu jj$ \\
decay mode at the CERN LHC
}

\author{K.~Iordanidis$^1$ and D.~Zeppenfeld}
\address{
Department of Physics, University of Wisconsin, Madison, WI 53706 \\[2mm]}
\address{and\\ [2mm]
$^1$Lincoln Capital Management Co., Suite 2100, 200 S. Wacker Dr., \\
Chicago, IL, 60606\\}
\maketitle
\begin{abstract}
The discovery of a heavy Higgs boson with mass up to $m_H = 1$ TeV at
the CERN LHC is possible in the $H\to W^+W^-\to l\nu jj$ decay mode. 
The weak boson scattering signal and backgrounds from $t\bar tjj$ and 
from $W+$jets production are analyzed with parton level Monte Carlo 
programs which are built on full tree level amplitudes for all subprocesses. 
The use of double jet tagging and the reconstruction of the W invariant mass
reduce the combined backgrounds to the same level as the Higgs signal.
A central mini-jet veto, which distinguishes the different gluon radiation 
patterns of the hard processes, further improves the signal to 
background ratio to about 2.5:1, with a signal cross section of 1~fb.
The jet energy asymmetry of the $W\to jj$ decay will give a clear signature
of the longitudinal polarization of the $W$s in the final event sample.
\end{abstract}
%
%\pacs{PACS numbers: 12.38.Bx, 13.60.-r, 13.87.Ce}
%
\newpage
%
%%%%%%%%%%%%%%%%%% MAIN TEXT %%%%%%%%%%%%%%%%%%%%%%%%%%%%%%%%%%%%%%%%%%%%%%%
%

\section{Introduction}

In the effort to determine the dynamics of the spontaneous breaking of the 
electroweak gauge symmetry, the discovery of the Higgs boson would be of 
prime importance. Detecting the Higgs boson is one of the biggest 
challenges for the CERN LHC~\cite{CMS,ATLAS}, 
both for a perturbative scenario for the symmetry 
breaking sector, with a Higgs boson mass below 
the $Z$-boson pair production threshold, and also if some strong 
interaction dynamics should be responsible for $SU(2)\times U(1)$ 
breaking~\cite{hunter,VLVL}. 
In both cases small signal rates, due to small usable decay 
branching fractions and/or small production rates, or large Standard 
Model (SM) backgrounds will have to be faced.  

In order to isolate a Higgs signal one will have to utilize all its 
characteristics. In turn this requires a simulation of the expected 
SM backgrounds with a high degree of detail, in a region of phase space 
where little or no experimental input exists at present. This problem is 
particularly acute for the search of a very heavy Higgs boson, with a 
mass above $\approx 600$~GeV. Here one will want to search for a Higgs
resonance in the scattering of longitudinal weak bosons, 
or, more generally, one will look for some structure in the invariant 
mass distribution of the produced weak boson pairs in electroweak processes 
of the type $q_1q_2\to q_3q_4V_LV_L$~\cite{VLVL}.  

Numerous studies over the past several years have indicated that for the
weak boson scattering signal to be identifiable, it is necessary to tag one or 
possibly two of the forward jets which arise from the scattered 
quarks~\cite{Cahn,Dtag,Gold,Froid,Stag}.
A second characteristic of the weak boson scattering process is the lack of 
color exchange between the two incident quarks, which distinguishes it 
from typical background processes which proceed via the $t$-channel 
exchange of color octet gluons. 
These different color structures are expected to lead to a rapidity gap 
signature for the signal, either in terms of soft hadrons, at low 
luminosity~\cite{troyan,bjgap} or in terms of mini-jets~\cite{bpz}. 

The small branching ratios of purely leptonic decays of the produced weak 
bosons can be overcome by studying the semi-leptonic modes, 
e.g. $H\to W^+W^-\to l\nu jj$. Here large backgrounds form $W$+jets 
production put a premium on good $W$-mass reconstruction of the two decay 
jets, in a situation where the large $W$ energy 
leads to a small separation of the two jets.
An advantage of this decay mode is the observability of the $W\to jj$ decay 
angular distribution which may allow a measurement of the longitudinal 
$W$ polarization of the signal.

Most of these points have been considered before. The detectability of the 
$H\to W^+W^-\to l\nu jj$ signal with jet tagging techniques, for example, 
has been discussed in the ATLAS and CMS technical design 
reports~\cite{CMS,ATLAS}.
However, these studies have been based on parton shower Monte Carlo 
programs and it is not clear how well these programs describe the 
high $p_T$ jets associated with the decaying $W$.  Also the color 
coherence effects which are at the basis of a rapidity gap trigger 
cannot be expected to be modeled correctly in these analyses.

In this paper we perform a complementary study, based on full QCD matrix 
elements of all subprocesses contributing to the signal and to the various 
backgrounds. We consider the signal process~\cite{Gold,Dicus,VBHiggs}
\bq
q_1q_2\to q_3q_4W^+W^- \to q_3q_4\; l\nu jj
\eq
(and crossing related ones) with a double forward jet tag on the two 
scattered quarks, $q_3$ and $q_4$. For the dominant $W$+jets QCD 
background we thus need the QCD matrix elements for all subprocesses 
leading to $W+4$~jet events~\cite{Berends,Madgraph}. Similarly, the 
potentially large $t\bar t\to bW^+\bar b W^-$ background needs to 
be simulated with two additional partons in the final state~\cite{Stange}, 
in order to account for the two tagging jets. When studying the consequences 
of different color structures on soft gluon radiation patterns, 
the ${\cal O}(\alpha_s)$ QCD corrections for the signal must be known as 
well~\cite{Duff}. While the parton level Monte Carlo programs for the 
individual subprocesses have been available in the literature, we here 
perform a first study of the $H\to W^+W^-\to l\nu jj$ 
mode with full QCD matrix elements for signal and background subprocesses. 

In Section~\ref{sec:2} we present these calculational tools in some detail. 
For the discussion of gluon radiation patterns we employ the truncated 
shower approximation which is briefly described at the end of that section. 
The isolation of the $H\to W^+W^-\to l\nu jj$ signal, with double 
forward jet tagging, but without considering the $W$-mass reconstruction 
from the $W\to jj$ decay is considered in Section~\ref{sec:3}. Here the 
hadronic system arising from the $W$-decay will be considered as a single 
jet. The properties of this $W$-decay jet, its internal dijet
structure, and the measurement of the $W$-mass is the subject of 
Section~\ref{sec:4}. Here we also consider the measurement of the $W$ 
polarization via the energy asymmetry of the two decay jets~\cite{Dtag}. 
Parameterizing the results of the $W$-mass analysis
in terms of a reconstruction efficiency, we return to the simpler analysis,
without simulating the $W\to jj$ dijet structure, in Section~\ref{sec:5}. We 
consider the mini-jet patterns which arise from additional gluon radiation 
in the Higgs signal and $W+4$~jets background,
or from $b$-quark jets in the $t\bar t$ background, as an additional selection 
criterion. With a central mini-jet veto above $p_{Tj} = 20$ GeV, the combined 
background is reduced well below the signal level, without significantly 
reducing the signal cross section. For an integrated luminosity of 
100 fb$^{-1}$, the expected event rate after all cuts is 99 (91) events 
for a $m_H=800$~GeV (1~TeV) Higgs boson signal, with a combined background 
of 41 events. These results suggest that the search for the Higgs boson 
at the CERN LHC can be
extended to the 1 TeV region, in the semi-leptonic Higgs decay channel.
Finally, a summary and our conclusions are given in Section~\ref{sec:6}.

\section{Cross Section Calculation for Signal and Background}\label{sec:2}

The signal process to be considered at lowest order is the subprocess
\begin{equation}\label{eq:signal}
      q_{1}q_{2} \rightarrow q_{3}q_{4}  
      W(\rightarrow l\nu) W(\rightarrow jj)
\end{equation}
and crossing related processes. In the following we require double forward 
jet tagging (of the jets corresponding to quarks $q_3$ and $q_4$)
and the presence of at least one additional high transverse momentum 
central jet (from $W\to jj$). These requirements are sufficient to eliminate 
soft and collinear divergences and they justify a few approximations in the 
cross section evaluations which will be discussed shortly. 

All cross section calculations are performed numerically, for $pp$ collisions
at a center of mass energy $\sqrt{s} = 14$~TeV. Individual 
subprocess cross sections are determined by numerically evaluating 
polarization amplitudes, mostly by using the amplitude techniques 
of Ref.~\cite{Helicity}. Even though this formalism is well suited to 
handle massive fermions, all quarks and $W$-decay leptons are treated 
in the massless approximation, except for the top-quarks. This approximation
greatly speeds up the calculations. Consistent with it, no 
Cabibbo-Kobayashi-Maskawa mixing is included in the calculation, even for 
incoming quarks. The error introduced by this approximation is well below 5\% 
and, hence, negligible compared to the typical uncertainties of a tree 
level calculation. $W$ decays are evaluated in the zero-width 
approximation. However, the $W\to f\bar f'$ decay amplitudes are fully 
implemented and, thus, all correlations between the decay fermions are 
included in our calculation. Finally, the phase space integrals are performed
with the adaptive Monte Carlo integration routine VEGAS~\cite{VEGAS}. 
The statistical error of all Monte Carlo integrals is below $1\%$, except 
for the $W+4$~jets process for which the statistical error on total cross
sections is $\approx 1.5\%$.

In all calculations, input parameters are a $Z$-mass of $m_Z=91.19$~GeV,  
$\sin^2 \theta_W=0.231$ for the weak mixing angle, and $\alpha=1/128.75$ 
for the QED fine-structure constant at the electroweak scale. 
From these $m_W=79.9$~GeV is derived at tree level. The 1-loop formula 
is used for the strong coupling constant $\alpha_{s}(\mu_R^{2})$, with 
$\alpha_{s}(M_{Z}^{2})=0.12$. For all processes, the MRS A 
parameterization of parton distribution functions is 
used~\cite{MRSA,pdflib}. Even though this parameterization is NLO 
and, hence, we are partially including higher order corrections, these 
ambiguities introduce negligible uncertainties. Finally, $b$-quark 
contributions to the initial state are neglected throughout.

In what follows, we give a brief account of calculational details for 
individual signal and background processes.

\subsection{The Electroweak Process $qq \rightarrow qq(g)W^{+}W^{-}$ }

\begin{figure}[t]
\epsfxsize=5.0in
\epsfysize=3.5in
\begin{center}
\hspace*{0in}
\epsffile{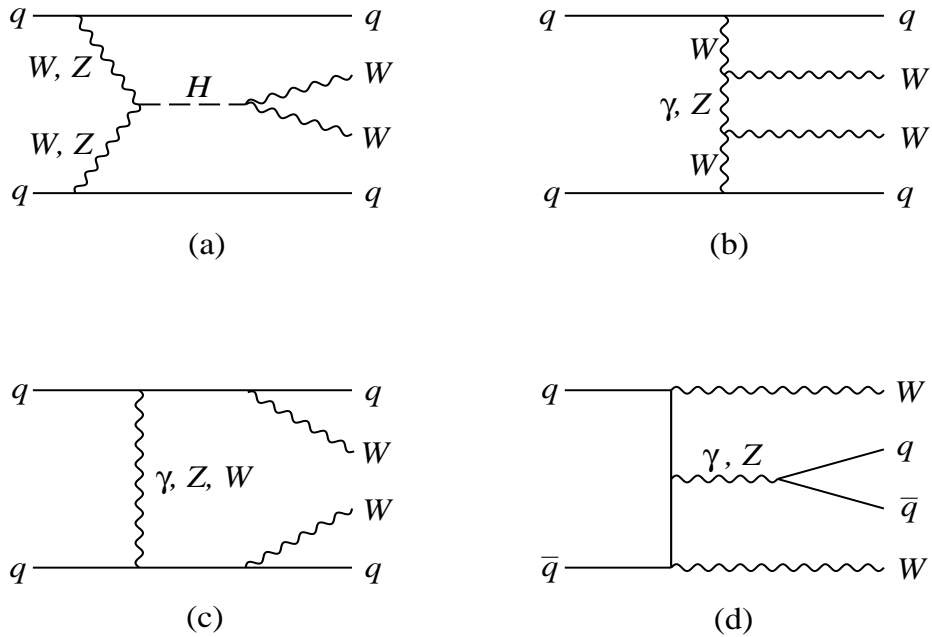}
\vspace*{0.2in}
\caption{
Feynman diagrams for the electroweak processes $qq\to qqW^+W^-$. 
Representative graphs are shown for (a) Higgs boson production via weak 
boson fusion, (b) generic weak boson scattering, (c) $W$ bremsstrahlung
off the quark lines, and (d) quark-antiquark annihilation.
\label{fig:feynsig}
}
\vspace*{-0.1in}
\end{center}
\end{figure}

The signal process at leading order is $qq\to qqH\to qqW^+W^-$ with 
subsequent $W$-decay, i.e. emission of the Higgs boson off a $t$-channel $W$ 
or $Z$ as shown in Fig.~\ref{fig:feynsig}(a). For a heavy Higgs boson 
mass ($m_H\agt 600$~GeV) the narrow Higgs width approximation is no longer 
applicable and all weak boson scattering processes (like the ones shown 
in Fig.~\ref{fig:feynsig}(b)) as well as $W$-bremsstrahlung off the quark 
lines (see Fig.~\ref{fig:feynsig}(c))
must be considered~\cite{Gold,Dicus,VBHiggs}. In principle we need to
evaluate the full set of ${\cal O}(\alpha_{ew}^{4})$ diagrams 
for a $W^{+}W^{-}$ final state, including contributions from $q\bar q$ 
annihilation graphs and fermion interchange graphs for identical quarks.
We will be requiring a double forward jet tag, however, which
puts the final state quarks into very different phase space regions 
and at large invariant mass. As a result, annihilation diagrams such as 
the one shown in Fig.~\ref{fig:feynsig}(d) and the interchange of identical 
fermions have very small contributions~\cite{Duff}. They will be neglected 
in the following. Within these approximations the helicity amplitudes for all
subprocesses are evaluated numerically, using the results of 
Ref.~\cite{VBHiggs}. 

The signal cross section, as discussed above, contains contributions
from non-resonant electroweak processes such as $W$ bremsstrahlung off the 
quark lines. Such contributions are independent of the mechanism for 
electroweak symmetry breaking and must be subtracted in order to get an
estimate of the Higgs boson signal cross section.
We model this continuum electroweak background by computing the
signal at $m_{H}=100~{\rm GeV}$. The actual signal cross section is then 
defined as the difference between the heavy Higgs and 
the $m_{H}=100~{\rm GeV}$ results, 
$\sigma_{sig}= \sigma(m_H)-\sigma(m_H=100~{\rm GeV})$.

In order to understand the characteristics of soft parton emission 
in the Higgs signal process, the ${\cal O} (\alpha_{ew}^{4} \alpha_{s})$ 
QCD corrections to the processes of Eq.~(\ref{eq:signal}) are needed.
The full set of real emission diagrams leading to a 
$W^{+}W^{-}+3~{\rm parton}$ final state was calculated in Ref.~\cite{Duff}
and we here use their results. The subprocesses to be considered are 
\begin{equation}
      q_{1}q_{2}  \rightarrow q_{3}q_{4} g\;  
      W(\rightarrow l\nu) W(\rightarrow jj)
\end{equation}  
and all crossing related processes like, for example,
\begin{equation}
      q_{1} g   \rightarrow \bar q_{2}q_{3}q_{4}   
      W(\rightarrow l\nu) W(\rightarrow jj)~.   
\end{equation}
Again, $s$-channel graphs corresponding to $q\bar q$-annihilation and
Pauli interchange graphs for identical quarks are neglected.  
For the Higgs signal calculation at leading order and at ${\cal O}(\alpha_s)$
both the renormalization and the factorization scales are 
set to the smallest transverse momentum of the final state partons.

\subsection{QCD $W+$~Jets Background}

In signal events with a high transverse momentum $W$ which decays 
hadronically, $W\to q\bar q$,  the two ``jets'' in the $W$ decay may 
merge and form a single high $p_T$ jet. In this case the signal events
produce a $W+3$~jets signature. The relevant QCD background for these
events comes from QCD processes with a $W$ and three jets in the 
final state. At leading order, two generic subprocesses contribute,
\begin{equation}
\begin{array}{lll}
    gg   & \rightarrow q_{1}\bar{q}_{2} g & W(\rightarrow l \nu) \\
    q_{1} \bar{q}_{2} & \rightarrow q_{3} \bar{q}_{4} g & 
                                          W(\rightarrow l \nu)\;. 
\end{array}
\end{equation}
We use the results of Refs.~\cite{Helicity,Barger_wjets} to calculate
the cross sections for these events. All crossing related processes are 
included in the calculation.

When investigating questions like the $W$-invariant mass resolution
in $W\to jj$ decays or the additional radiation of soft partons in 
$qq\to qqW^+W^-$ events, the QCD $W+$~jets backgrounds with four partons 
in the final state are needed. 
The subprocesses that contribute can be classified as 
6 quark processes, 4 quark plus 2 gluon processes, and 2 quark plus 4 gluon
processes,
\begin{equation}
\begin{array}{lll}
    q_{1} \bar{q}_{2} & \rightarrow q_{3} \bar{q}_{4} q_{5} \bar{q}_{6}  & 
          W(\rightarrow l \nu) \\
    q_{1} \bar{q}_{2} & \rightarrow q_{3} \bar{q}_{4} g g   & 
          W(\rightarrow l \nu) \\
    gg   & \rightarrow q_{1}\bar{q}_{2} g g & W(\rightarrow l \nu) \;.
\end{array}
\end{equation}
The cross sections for these and all crossing related subprocesses were 
first calculated in Ref.~\cite{Berends}. Here we use equivalent 
matrix elements which were computed by generating the helicity 
amplitudes with the program MadGraph~\cite{Madgraph}.

For the $W+n$~jets QCD processes the factorization scale
is set equal to the smallest transverse momentum of the final state partons.
At leading order these cross sections are proportional to $\alpha_s^n$, and
the strong coupling constant $\alpha_{s}$ is evaluated at the corresponding 
transverse momentum of each final state jet, i.e., 
$\alpha_{s}^{n} = \prod_{i=1}^{n} \alpha_{s}(p_{T,{\rm jet}_{i}})$.

\subsection{ $t\bar tjj$  Background}

For the $t\bar{t}$ background, the b-quarks from the $t \to W b$ decay are
produced mainly in the central region, with the two forward jets resulting 
mainly from QCD radiation. The relevant leading order process is the 
production of $t\bar{t}$ pairs in association with two jets, which 
includes the following subprocesses
\begin{equation}
\begin{array}{llll}
    gg          & \rightarrow t\bar{t} gg & 
                  \rightarrow W^{+} b W^{-} \bar{b} & g g  \\
    q\bar{q}    & \rightarrow t\bar{t} gg & 
                  \rightarrow W^{+} b W^{-} \bar{b} & g g \\
    q_{1} q_{2} & \rightarrow t\bar{t} q_{1} q_{2} & 
                  \rightarrow W^{+} b W^{-} \bar{b} & q_{1} q_{2}. 
\end{array}
\end{equation}
The exact matrix elements for the ${\cal O} (\alpha_{s}^{4})$ processes
are evaluated, including all the crossing related subprocesses. 
The Pauli interference terms between identical quark flavors in the process 
$q_{1}q_{2}\rightarrow t\bar{t}q_{1}q_{2}$ are neglected, with little effect 
in the overall cross section rate, due to the 
large differences in the transverse momenta and energies of the final state 
partons~\cite{Stange}. 
The top quark decays are simulated in the narrow width 
approximation, and its mass is set to  $m_{t}=175$ GeV. 
The structure function scale is chosen to be the smallest transverse energy
of the final state partons before the top quark decay. 
The strong coupling constant $\alpha_{s}$
is evaluated at the corresponding transverse energy of the final 
state partons, prior to the top quark decay, i.e.,
$\alpha_{s}^{4} = \alpha_{s}(E_{T}(t)) \alpha_{s}(E_{T}(\bar{t}))
                 \alpha_{s}(p_{T,{\rm jet}_{1}}) 
                 \alpha_{s}(p_{T,{\rm jet}_{2}})$.

In order to study the effects of additional parton radiation
in the top quark background, one would like to evaluate the $t\bar{t}+3$~jets 
cross sections as well. Since such a calculation is not available yet,
we only consider the additional central jet activity arising from the 
$b$-quarks which are associated with the top-quark decays. 
The probability of a $b$-quark with $p_{T}^{b}>20$~GeV to be 
identified as one of the two forward tagging jets was found to be small 
($\approx 6\%$)~\cite{Mthesis}. With the transverse momentum and separation 
requirements on the two tagging jets to be discussed below, only this small
fraction of the $t\bar t jj$ background is affected by collinear and infrared 
singularities. Instead of dropping these events altogether we regularize the
singularities with the truncated shower approximation (TSA).

\subsection{The Truncated Shower Approximation}\label{sec:2d}

As the transverse momentum of the softest parton becomes small, 
the perturbative calculation of the ${\cal O}(\alpha_{s})$ cross section 
for both signal and background breaks down due to the collinear and 
infrared divergences associated with gluon emission. In a complete 
next-to-leading order (NLO) calculation these divergences are 
eliminated due to the cancellation between virtual and real emission 
corrections. For the multi-parton processes considered here, a full 
NLO treatment is not yet possible, however. 
Instead, we model the effects of multiple soft gluon emission by using
the truncated shower approximation (TSA)~\cite{TSA}. The TSA correctly 
reproduces the normalization of the lowest order cross section (which 
is free of infrared and collinear divergences) and it agrees with the full 
NLO calculation when the emission of one additional hard parton is 
considered. At the same time the TSA provides a model for the collective 
effects of multiple soft parton emission in events with $n$ hard jets. 
Specifically, the tree level $n+1$-jet cross section is replaced by
\begin{equation}
\label{TSA}
      \sigma(n+1~ j)_{TSA} = K~
      \int{ f_{TSA}(p_{Tj,min}) 
            { d\sigma(n+1~ j)_{TL} \over   
              d p_{Tj,min} }  d p_{Tj,min} }\; .
\end{equation}
Here  $p_{Tj,min}$ is the transverse momentum of the softest jet, 
\begin{equation}
\label{TSA_fac}
      f_{TSA}(p_{Tj,min}) = 1 - \exp \left( {- { p_{Tj,min}^{2} \over
      p_{TSA}^{2} } } \right)
\end{equation}
is a Gaussian cutoff factor, and $K$ is a multiplicative factor that 
effectively includes the full 1-loop corrections. It has been shown 
that the $K$-factor for vector boson scattering in $pp$ collisions 
is small ($K=1.06$ at the CERN LHC for 
$m_{H}=800$~GeV)~\cite{Kfactor}. Since $K$-factors are unknown for the 
background processes we set $K=1$ throughout this study.
The parameter $p_{TSA}$ is chosen so that the cross section of 
Eq.~(\ref{TSA}) correctly reproduces the lower order $n$ jet 
cross section.

For jet transverse momenta $p_{Tj}<p_{TSA}$, the TSA leads to a reduction
of the transverse momentum distribution of the hard $n$-jet system
which simulates the canceling of multiple soft parton momenta.
Thus $p_{TSA}$ provides an estimate of the jet transverse momentum
scale below which the emission of multiple soft gluons becomes 
important. In the phase space regions for hard jets to be discussed
below, we find values of order $p_{TSA}\approx 40$~GeV for the $W+4$~jets
QCD background as compared to $p_{TSA}\approx 8$~GeV for the signal.

For the $t\bar t$ background we apply the TSA only to those events 
where one of the $b$-quarks arising from the top quark decays gives
rise to at least one of the two forward tagging jets. In such events
one of the two additional final state partons can be soft, and the cross 
section is enhanced in the region of the phase space where the 
transverse momentum of that jet is small. In order to avoid this 
singular behavior, the TSA is applied only to this fraction of 
events, with $p_{TSA}^{t\bar{t}} = 42$~GeV~\cite{bpz}.

\section{Deciphering the Higgs Signal from Background}\label{sec:3}

The signal process $qq\to qqH\to qq\,(W\to l\nu)\,(W\to q\bar q)$ gives 
rise to two forward tagging jets, one (or two) hard central jets from the
hadronic $W$ decay, and a leptonic $W$ decay signature.
In order for the $W$ decay products to be identified, it is required that 
each event contains a charged lepton, $l$ (either $e$ or $\mu$), in the 
central region, with
\begin{equation}
\label{lepton_cut}
      p_{Tl} > 100~{\rm GeV}\, ,  \qquad
      |\eta_{l}| < 2\, , \qquad
      \Delta R_{lj} = \sqrt{(\eta_{l}-\eta_{j})^{2} + (\phi_{l}-\phi_{j})^{2}}
      > 0.7\, ,
\end{equation}
where $p_{Tl}$ is the transverse momentum of the lepton, $\eta_{l}$ is the 
lepton pseudo-rapidity, and $\Delta R_{lj}$ is the distance between the 
lepton and any identified jet in the pseudo-rapidity-azimuthal angle plane.
In addition, it is assumed that each event has large missing transverse 
momentum due to the neutrino of the $W\rightarrow l\nu$ decay,
\begin{equation}
\label{misspt_cut}
      \pslasht > 100~{\rm GeV}\; .
\end{equation}
All final state partons are identified as jets if they satisfy
\bq\label{eq:jetdef}
p_{Tj} > 20~{\rm GeV}\;, \qquad  |\eta_j|<4.5\; ,
\eq
and if they are well separated in the 
pseudo-rapidity-azimuthal angle plane, with 
\bq
\Delta R_{jj} = \sqrt{(\Delta \eta)^{2} + (\Delta \phi)^{2}} > 0.7\; .
\eq
The requirements of Eq.~(\ref{eq:jetdef}) are superseded by more stringent 
requirements for the tagging jets and for the Higgs decay products. The 
hadronically decaying $W$ of the Higgs boson signal is identified by requiring 
the existence of a large transverse momentum jet in the central region,
\begin{equation}
\label{jc_cut}
      p_{Tj}^{c} > 300~{\rm GeV}\;, \qquad \qquad |\eta_{j}^{c}| < 2\; .
\end{equation}

\begin{figure}[t]
\vspace*{0.5in}            
\begin{picture}(0,0)(0,0)
\includegraphics{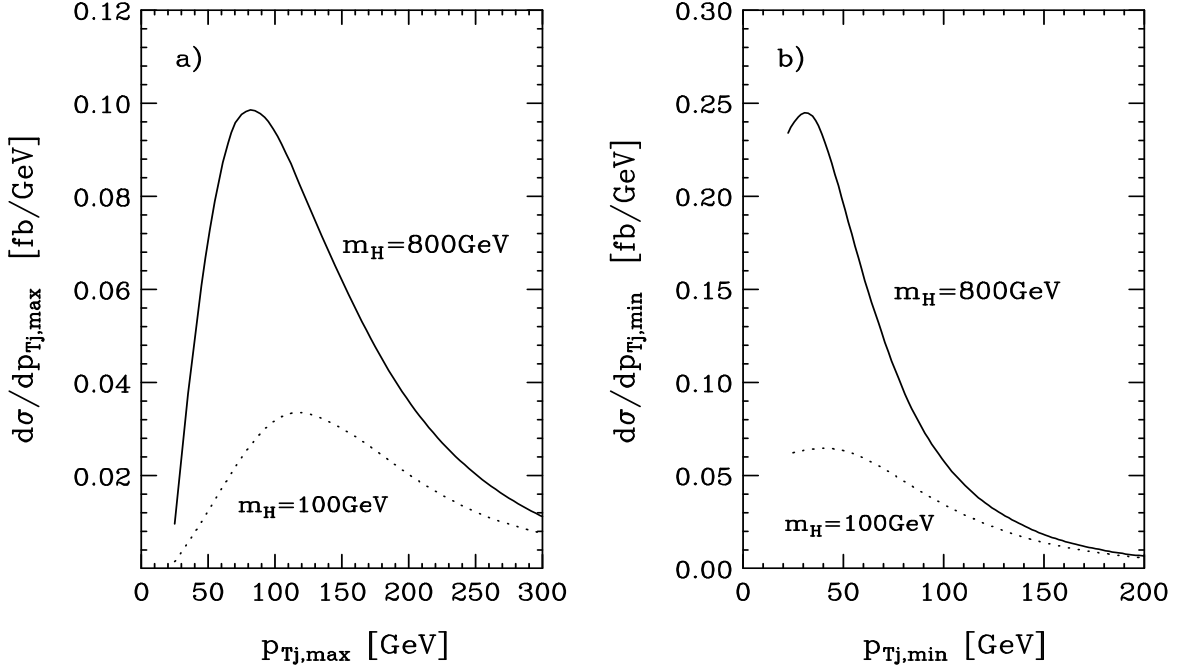}
\end{picture}
\vspace{8.5cm}
\caption{
Transverse momentum distributions (a) for the highest 
and (b) for the lowest $p_{T}$ forward jets at $\protect\sqrt{s}= 14$~TeV. 
For each event, a high transverse momentum
lepton in the central region is required, $p_{Tl}>100$~GeV and 
$|\eta_{l}|<2$, as well as missing transverse momentum of $\pslasht>100$~GeV.
In addition a minimum of three visible jets are required, each 
with $p_{T}>20$~GeV.
The solid line represents the full signal calculation for $m_{H}=800$~GeV, 
while the dotted line represents the continuum electroweak background 
($m_{H}=100$~GeV).     
\label{fsignalpt}
}
\vspace{2mm}
\end{figure}

The two quark jets in the process $qq\to qqH$ are tagged by requiring the 
presence of two additional jets, in the forward and backward regions, with
\begin{equation}
\label{tag1_cut}
      p_{Tj}^{tag1} > 50~{\rm GeV}\; , \qquad 
      2 < \left| \eta_{j}^{tag1} \right| < 4.5\; ,
\end{equation}
and 
\begin{equation}
\label{tag2_cut}
\begin{array}{ll}
      p_{Tj}^{tag2} > 30~{\rm GeV}\; , \qquad  & 
      \left\{ \begin{array}{rl}
  -4.5 <  \eta_{j}^{tag2} < -2   & \qquad \mbox{if $\eta_{j}^{tag1}>0$} \\
   2   <  \eta_{j}^{tag2} < 4.5  & \qquad \mbox{if $\eta_{j}^{tag1}<0$}
              \end{array} \right. .
\end{array}
\end{equation}
The asymmetric transverse momentum requirement on the two tagging jets is
motivated by the fact that one of the jets has substantially higher median
$p_T$ than the other, as shown in Fig.~\ref{fsignalpt}.

The resulting cross sections for the signal and the background are shown 
in the first column of Table~\ref{Table_tag}. 
The $W$+3 jets background is a factor of 20 larger than the signal, whereas 
the $t\bar{t}$ background is a factor of three larger. 
In contrast, the electroweak continuum background is strongly 
suppressed by double tagging, due to the fact that the rapidity 
distribution of the two tagging jets for the electroweak background peaks
in the central region~\cite{Uliletter}.
At this level it contributes only $\sim 10\%$ to the signal cross section.

\begin{table}[htb]
\caption{Signal and background cross sections $B\sigma$ in fb after
double jet tagging. The decay lepton acceptance requirements are 
$p_{T_{l}}>100$ GeV and $|\eta_{l}|<2$, and each event is required to have 
missing transverse momentum $\pslasht>100$ GeV. The signal is defined as 
$\sigma(m_H)-\sigma(m_H=100$~GeV). }
\label{Table_tag}
\vskip 0.15in
\begin{tabular}{lccc}
         & Double jet         & +jet energy    & + lepton-tagging \\ 
         &    tagging         &     cut        & jet separation   \\
         & [Eq.(\protect\ref{jc_cut})-(\protect\ref{tag2_cut})]   
         &   [Eq.(\protect\ref{energy_cut})]
         & [Eq.(\protect\ref{lepton_sep})]   \\ \hline
$m_{H}=800$ GeV  &  3.15   &   1.96  &  1.58 \\
$m_{H}=100$ GeV  &  0.26   &   0.18  &  0.10 \\
$W$+3 jets       &  66.3   &   18.2  &  8.36 \\
$t\bar{t}+jj$    &  8.01   &   3.05  &  1.55 \\[0.2in]
\underline{signal}:  &  &  &\\
$m_{H}=800$ GeV  &  2.89  & 1.78  &  1.48 \\
\end{tabular}
\end{table}

For the $W$+3 jets cross section with the tagging requirements of 
Eqs.~(\ref{tag1_cut}) and~(\ref{tag2_cut}), it is important to ensure 
that it is a well defined hard scattering process for which a perturbative
evaluation is reasonable.
In order to investigate the effect of double tagging on the cross section,
we computed the $W$+1 jet cross section with the 
single jet satisfying Eq.~(\ref{jc_cut}), and the $W$+2 jets cross section
with the two jets satisfying Eqs.~(\ref{jc_cut}) and~(\ref{tag1_cut}).
The $W$-decay leptons must satisfy the cuts of 
Eqs.~(\ref{lepton_cut},\ref{misspt_cut}). 
The $W$+1 jet cross section is 2.16 pb, whereas 
the $W$+2 jets cross section is 0.57 pb. The corresponding reduction factors 
are 3.8 from $W+1$~jet to $W+2$~jets and 8.7 from $W$+2 jets to $W$+3 jets,
respectively. These factors are typical for perturbative QCD processes
with successively larger numbers of jets and lend credence to the use
of perturbation theory in the evaluation of the QCD $W+n$~jets
backgrounds.

The signal to background ratio can be further improved by utilizing 
differences in the tagging jet characteristics between signal 
and background. The two forward jets for the signal are very energetic and 
their energy distributions decline slower than the energy distributions of 
the two forward jets for the $W$+3 jets and the $t\bar{t}jj$ 
backgrounds~\cite{Froid,Stag,Mthesis}.
The softer jet energy distributions for the background reflect the fact 
that these jets tend to come from soft gluon radiation in the forward region.
By requiring that both tagging jets satisfy
\begin{equation}
\label{energy_cut}
      E_{j}^{tag1,2} > 500~{\rm GeV}\; ,
\end{equation}
the signal to background ratio can be improved by more than a factor of two
(see second column of Table~\ref{Table_tag}).
A second distinction arises in the pseudo-rapidity separation of the charged 
decay lepton and the closest tagging jet. For the Higgs signal there is little 
correlation between the two because the leptonic $W$-decay arises from the 
decay of a scalar particle which moves slowly in the laboratory frame. 
By contrast the $W+3$ jets background contains many events with $W$ 
bremsstrahlung off one of the tagging jets, and such events favor a 
small separation between the jet and the decay lepton. These differences 
are exploited by imposing a cut,
\begin{equation}
\label{lepton_sep}
      |\eta_{j}^{tag1,2}-\eta_{l}| > 2\; ,
\end{equation}
on the separation between the decay lepton and the two tagging jets.
The signal and background cross sections after all hard cuts are
shown in the third column of Table~\ref{Table_tag}. The $W$+3 jets 
background is still a factor of 6 larger than the signal, whereas 
the $t\bar{t}$ background has been reduced to the same level as the signal.

We find that any further hardening of the acceptance criteria discussed
so far will degrade the signal rate appreciably, with only marginal 
improvement to the signal's statistical significance. 
Additional information is needed in order to further
suppress the background without significantly degrading the signal 
cross section. This is the focus of the following two sections.

\section{Reconstruction of the $W\to jj$ Decay}\label{sec:4}

One additional piece of information is provided by the internal structure of 
the large transverse momentum jet in the central region which represents the
hadronically decaying $W$ of the Higgs signal. The invariant mass of this
system, which may or may not be resolvable into two separate jets, provides
an important criterion for suppressing the QCD $W$+jets background. Whenever 
the pair of jets from the hadronic $W$ decay can be resolved, further 
information is gained.  A Higgs boson decays mostly into longitudinally 
polarized $W$s whereas backgrounds with real $W$s are dominated by 
transversely polarized weak bosons. The angular distribution and the 
energy asymmetry of the two central jets are sensitive to the polarization 
of the $W$ boson, and can be used in order to test 
whether or not the reconstructed $W$ is the longitudinally polarized
decay product of the Higgs boson~\cite{Dtag,Field}.

Some of these questions have been studied previously with the aid of 
parton shower Monte Carlo programs~\cite{CMS,ATLAS,Field,CMSnote}. Since 
we have a full QCD matrix element calculation available for the production 
of $W+4$~jet events, we can avoid the approximations inherent in a parton 
shower program and use full tree level QCD to simulate the two forward 
jets as well as the two central jets which would fake the hadronically 
decaying $W$. The $t\bar{t}$ background is not included in 
this study of jet mass effects since, similar to the signal process,
the observed central jet pair is the result of the decay of a real $W$ 
boson, which, typically, is longitudinally polarized.

When using  $W$ mass reconstruction, the experimental resolution of the dijet
mass is the limiting factor. In order to model these experimental errors, the 
lateral granularity of the detector must be taken into account. Following 
the design specifications of the CMS detector~\cite{CMSletter}, 
we divide the legoplot into cells of size
\begin{equation}
\label{cells}
      \Delta\eta \Delta\phi = 0.1 \times 0.1~ .
\end{equation}
The momentum vectors of the two central jets are then corrected to point to 
the center of the cell. This correction is applied to the Higgs signal 
and to the continuum electroweak background, but not to the 
$W$+4 jets background. For the former these smearing effects are important,
due to the resonance in the dijet invariant mass spectrum at $m_{jj}=m_W$, 
while the background exhibits a fairly flat dijet mass spectrum which 
mitigates any smearing corrections. 
The finite energy resolution of electromagnetic and hadronic calorimeters 
affects both signal and background cross sections because energy and 
transverse momentum distributions are typically quite steep. 
These energy resolution effects are taken into account by Gaussian 
smearing of the overall energy 
scales of massless parton four-momenta, with relative energy 
uncertainties~\cite{CMSletter},
\begin{equation}
\label{smear1}
      {\Delta E_{em} \over E} =  {0.03 \over \sqrt{E}} \otimes
                                 {0.15 \over E} \otimes
                                  0.005 \; ,
\end{equation}
and
\begin{equation}
\label{smear2}
      {\Delta E_{had} \over E} = \left\{ 
      \begin{array}{ll}
            \frac{\textstyle 0.8}{\textstyle \sqrt{E}} \otimes
            {\textstyle 1.0 \over \textstyle E} \otimes 0.03 & 
             \mbox{\qquad when $|\eta_{j}|\leq 2.5$} \\
            {\textstyle 1.0 \over \textstyle \sqrt{E}} \otimes
            \frac{\textstyle 3.0}{\textstyle E} \otimes 0.05 & 
             \mbox{\qquad when $2.5<|\eta_{j}|<4.5$} 
      \end{array}
            \right . 
\end{equation}

Here '$\otimes$' means that the terms are added in quadrature.
Energy smearing according to Eqs.~(\ref{smear1},\ref{smear2}) is applied
to both the signal and the $W+4$~jets background.

\subsection{Resolution of Jet Pairs from $W\rightarrow q\bar{q}^{'}$ Decay}

The resolution of the two jets from $W\rightarrow q\bar{q}^{'}$ decay
depends on the angular separation,
\bq
\Delta R_{jj}^c= \sqrt{(\eta_{j_1}-\eta_{j_2})^2+(\phi_{j_1}-\phi_{j_2})^2}\;,
\eq
of the two partons in the legoplot. Furthermore, in order to suppress the 
QCD $W+4$~jets
background, it is advantageous to raise the transverse momentum threshold 
for each of the two central jets above our nominal value of 20~GeV. 
For the study of $W$ hadronic decay we thus require the existence
of two central jets, with the following acceptance requirements
which are added to the requirements of the previous section:

\begin{enumerate}
\item Each of the central jet candidates must have large transverse 
      momentum and be in the central rapidity region,
     \vspace*{-0.1in}
     \begin{equation}
     \label{jet_cut_wjj}
        p_{Tj}^{c} > 50~{\rm GeV}, \qquad |\eta_{j}^{c}| < 2~.
     \end{equation}
     All jets passing this criterion form candidate pairs for 
     the hadronic $W$ decay products.
\item For each candidate pair, the reconstructed $W$ must have
      large transverse momentum, and it must lie in the 
      hemisphere opposite to the lepton-neutrino pair, 
\begin{equation}
     \label{jj_cut_wjj}
    p_{T}^{jj} > 300~{\rm GeV}, \qquad |\phi_{jj}-\phi_{l\nu}| > 90^{\circ},
\end{equation}
where $\phi_{jj}-\phi_{l\nu}$ is the azimuthal angle between the jet-jet pair
and the lepton-neutrino pair. 
\item Finally, it is required that the two central jet candidates are
separated by 
\begin{equation}
   \label{rjjc_min_cut}
   0.2 < \Delta R_{jj}^{c} < 1.0~.     
\end{equation}
\end{enumerate}
%If more than one jet pair meets the above requirements, then the jet pair
%with the largest total transverse energy is selected.

\begin{figure}[thb]
\vspace*{0.2in}            
\begin{picture}(0,0)(0,0)
\includegraphics{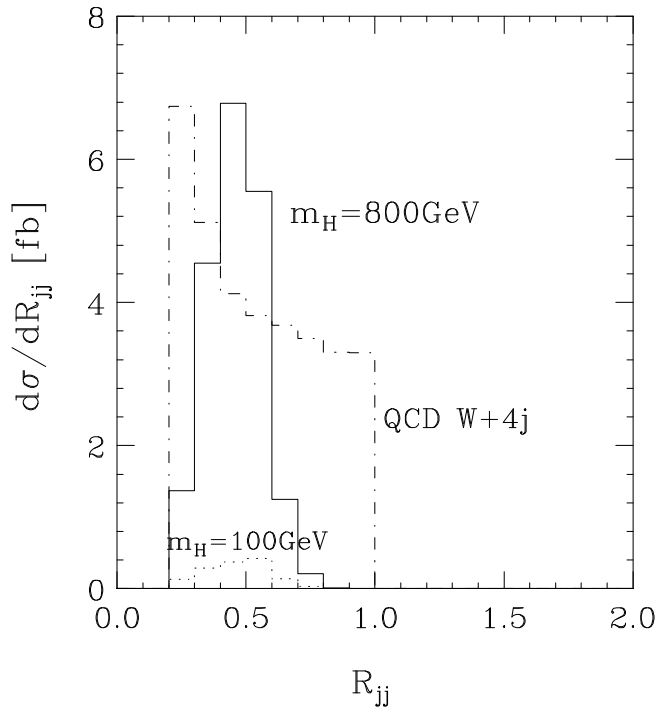}
\end{picture}
\vspace{9.3cm}
\caption{ 
Separation between the two central jets in the pseudo-rapidity-azimuthal 
angle plane. % at $\protect\sqrt{s} = 14$ TeV.  
The solid histogram represents the full signal calculation for 
$m_{H}=800$ GeV, which still contains the continuum electroweak background 
(dotted line), calculated in terms of the $m_{H}=100$ GeV cross section.
The QCD $W$+4 jets background is given by the dash-dotted histogram.
Energy smearing according to 
Eqs.~(\protect\ref{smear1},\protect\ref{smear2}) is applied to signal 
and background processes. Finite detector granularity (see 
Eq.~(\protect\ref{cells})) is taken into account for the electroweak 
processes.
\label{fig:drjj}
}
\vspace{5mm}
\end{figure}

The $\Delta R_{jj}^c$ distributions for the signal and the background are 
shown in Figure~\ref{fig:drjj}. The minimum $\Delta R_{jj}^c$ 
requirement of $0.2$ is still sufficient to eliminate the final state 
collinear singularity of the $W$+4 jets cross section at  
$\Delta R_{jj}^c\rightarrow 0$. Notice also that the maximum separation cut,
$\Delta R_{jj}^c<1$, has an appreciable effect on the $W+4$ jets cross section
only. Due to the large transverse momentum carried by the $W$ 
(see Eq.~(\ref{jj_cut_wjj}))
and the associated strong boost of the $W$ decay products, the two quarks 
from the $W$ decay are rarely separated by more than $\Delta R_{jj}^c=0.7$ in
the laboratory frame. 

\begin{table}[thb]
\caption{ Signal and background reduction factors resulting from an 
     analysis of the central cluster which is a candidate for hadronic
     $W$ decay. The first column gives the efficiency of reconstructing
     two jets in the central cluster, within the cuts of 
     Eqs.~(\protect\ref{jet_cut_wjj}--\protect\ref{rjjc_min_cut}). Requiring
     the invariant mass of these two jets to lie in the $M_W\pm 15$~GeV
     window yields the additional reduction factor of the second column.
     The product of the two yields the total efficiency listed in column three.
        }
\label{Table_Wjj}
\vskip 0.15in
\begin{tabular}{lccc} 
         & $ { \sigma( W \rightarrow {\rm jj}) \over 
               \sigma( W \rightarrow 1 {\rm jet})     }$  & 
           $ { \sigma( W \rightarrow {\rm jj},M_{jj} cut) \over
               \sigma( W \rightarrow {\rm jj}) }$ & 
           Efficiency\\ 
         &   [Eq.~(\ref{cells})--(\ref{rjjc_min_cut})]   
         & [Eq.~(\ref{wmass_cut})] &  \\ \hline
$m_{H}=800$ GeV  &  0.87  & 0.83  &  0.71 \\
$m_{H}=100$ GeV  &  0.67  & 0.86  &  0.57 \\
$W$+jets         &  0.27  & 0.21  &  0.055 \\[0.2in]
\underline{signal}:  &  &  & \\
$m_{H}=800$ GeV  &  0.88  & 0.83  &  0.73 \\
\end{tabular}
\end{table}

In the previous Section, no requirement was imposed on the internal structure
of the central hard jet. Resolving it into two jets, 
corresponding to the $W\to q\bar q'$ decay, will lead to a reduced rate 
for the signal. The corresponding cross section reduction factors, after 
the cuts of Eqs.~(\ref{jet_cut_wjj}--\ref{rjjc_min_cut}),
are listed in the first column of Table~\ref{Table_Wjj}. For the electroweak
processes, the effects of detector granularity and of energy smearing must be
included also in the determination of the single jet cross sections which 
serve to normalize these reduction factors.
For the signal cross section, $86.4\%$ of the events pass the selection 
criteria. The reduction in the cross section 
comes mostly from the minimum $p_{T}$ requirement of 50 GeV for each of the 
two central jets. Only two thirds of the electroweak continuum 
background survive the cuts.

The continuum $W+$~jets background is reduced by approximately a factor 4
when the resolution of the central jet into two hard jets of $p_T>50$~GeV
is required. This reduction is gratifying since it indicates that the use 
of perturbative QCD is still warranted, in spite of the small minimal 
separation of 0.2 for the two almost collinear partons which mimic the
$W\to jj$ decay.

\subsection{Reconstruction of the $W$ Invariant Mass}

\begin{figure}[t]
\vspace*{0.5in}            
\begin{picture}(0,0)(0,0)
\includegraphics{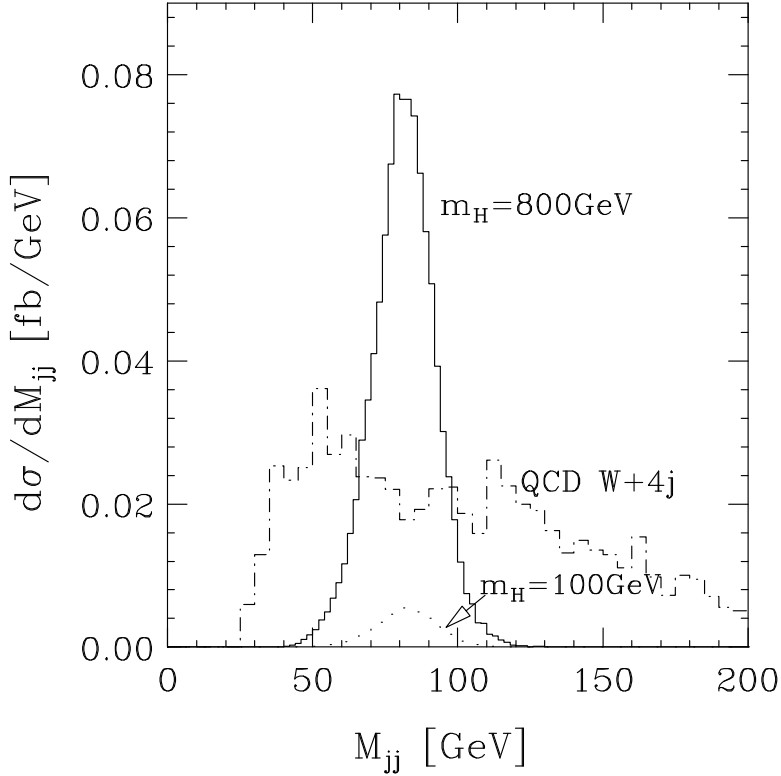}
\end{picture}
\vspace{9.3cm}
\caption{
Reconstructed dijet invariant mass distribution of the 
hadronically decaying $W$.  %, at $\protect\sqrt{s} = 14$ TeV. 
The solid histogram represents the full signal calculation for 
$m_{H}=800$ GeV, with the continuum electroweak background ($m_{H}=100$ GeV)
given by the dotted line. 
The QCD $W$+4 jets background is given by the dash-dotted histogram.
Finite detector resolution is taken into account as in Fig.~\ref{fig:drjj}.
\label{figwmass}
}
\vspace{5mm}
\end{figure}

A further reduction of the background is achieved by requiring that the 
two hard central jets are consistent in invariant mass with a hadronically 
decaying $W$-boson. The reconstructed invariant mass distribution of these 
two jets is shown in Fig.~\ref{figwmass}.

The distributions for the signal and the continuum background are  
narrow and peak at $M_{W}$. The distribution for the $W$+4 jets 
background, on the other hand, is flat, reflecting the fact that in 
this case the jet pair is not the decay product of a real $W$ boson.
These differences are exploited by a simple invariant mass cut on the central
dijet pair,
\begin{equation}
\label{wmass_cut}
      M_{W} - 15~{\rm GeV} < M_{jj}^{c} < M_{W} + 15~{\rm GeV}\;.
\end{equation}
The chosen mass window of $\pm 15$~GeV is motivated by the signal width in 
Fig.~\ref{figwmass} and agrees with results of a more complete detector 
simulation~\cite{CMS,ATLAS}. The reduction factors for both the signal and 
the background, due to the dijet mass cut of Eq.~(\ref{wmass_cut}), are 
given in the second column of Table~\ref{Table_Wjj}. 
The $W$+4 jets background is reduced by an additional factor of 5,
whereas $83\%$ of the signal events survive the cut.
The overall efficiency of the central jet resolution and the $W$ invariant 
mass cut is given in the third column of Table~\ref{Table_Wjj}.
For the Higgs boson signal, $73\%$ of the events survive all cuts,
in contrast to only $5.5\%$ of the events for the $W$+4 jets background. 
At this level, the $W$+4 jets background is a factor of 2.3 smaller than 
the signal. Even if the central jet pair cannot be resolved, it may 
still be possible to measure the invariant mass of the broad central 
jet representing the hadronically decaying $W$ boson. The reduction 
factors in the second column of Table~\ref{Table_Wjj} and the cross 
section values in the last column of Table~\ref{Table_tag} indicate that 
the $W$ invariant mass cut would reduce the $W$+ jets background to 
the level of the signal.

\subsection{Measurement of the $W$ Polarization}

\begin{figure}[t]
\vspace*{0.5in}            
\begin{picture}(0,0)(0,0)
\includegraphics{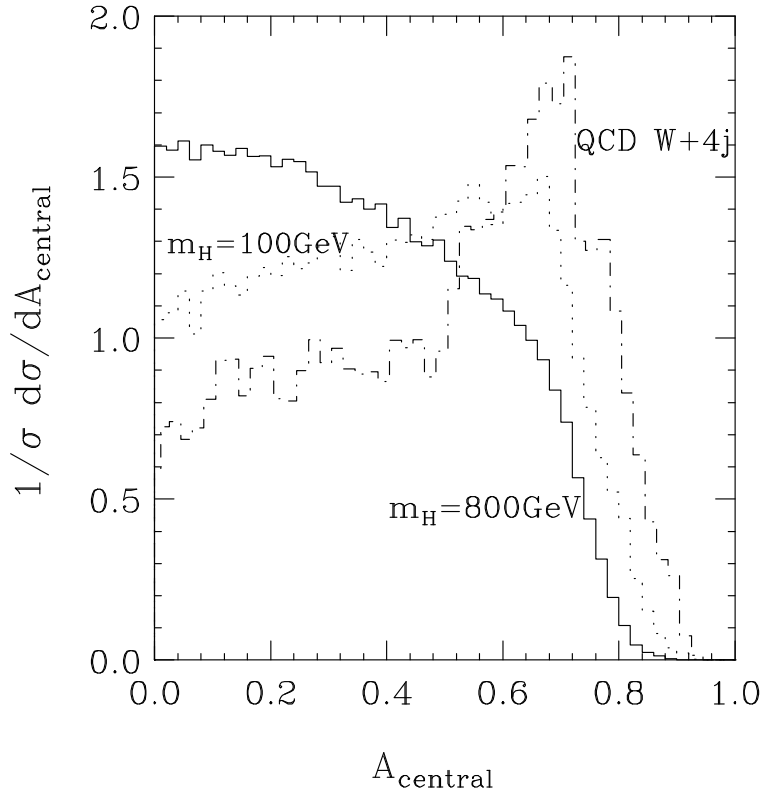}
\end{picture}
\vspace{9.3cm}
\caption{
Energy asymmetry distribution of the two central jets, which are identified 
as the $W\to jj$ decay products. % at $\protect\sqrt{s} = 14$ TeV. 
No dijet invariant mass cut is imposed. The solid histogram represents 
the full signal calculation for $m_{H}=800$~GeV, while the continuum 
electroweak background ($m_{H}=100$~GeV) is given by the dotted line. 
The dash-dotted histogram represents the QCD $W$+4 jets background.
Finite detector resolution is taken into account as in Fig.~\ref{fig:drjj}.
\label{figasym} 
}
\vspace{5mm}
\end{figure}

Any polarization of the hadronically decaying $W$ affects the 
angular distributions of the two $W$ decay jets. A transversely polarized 
$W$ yields a $1+{\rm cos}^2\theta^*$ distribution whereas longitudinally
polarized $W$s produce a ${\rm sin}^2\theta^*$ distribution. Here $\theta^*$
is the polar angle of one of the decay jets with respect to the $W$ direction,
in the $W$ rest frame. The approximate alignment of the thrust axis with 
the $W$ direction for transverse $W$s produces two jets of quite different 
energies after boosting into the laboratory frame. Longitudinally polarized
$W$s, on the other hand lead to approximately equal jet energies. 
The energy asymmetry, $A$, of the two central jets therefore
is an excellent variable to confirm the longitudinal polarization of the $W$s
expected in Higgs boson decay~\cite{Dtag}. It is defined as
\begin{equation}
   A = {\left| E_{1} - E_{2} \right | \over
               E_{1} + E_{2} }
\end{equation}
where $E_{1}$ and $E_{2}$ are the energies of the two central jets in the 
laboratory frame.
The energy asymmetry distributions for the $m_H=800$~GeV signal and for the
electroweak and the $W+4$~jet QCD backgrounds are shown in Fig.~\ref{figasym},
without imposing the dijet invariant mass cut of Eq.~(\ref{wmass_cut}). 
Each distribution is normalized to the corresponding integrated cross section.
The difference between the distributions for the signal and the background is
striking. For the longitudinally polarized signal, the two jets have very 
similar energies so the distribution peaks at $A=0$. In contrast, the two 
central jets of the $W+4$~jets background and the two jets arising 
from the decay of a transversely polarized $W$ in the electroweak continuum 
background have substantially 
different energies, with the distribution peaking at large values of $A$.

Clearly, an energy asymmetry cut ({\em e.g.} at $A=0.5$) would further 
improve the signal to background ratio (by a factor of about 1.6). 
We do not impose such a cut here
because there are other tools, namely a jet veto on the additional 
minijet activity in the central region, which can be exploited for an 
adequate background rejection. The energy asymmetry can then be used 
to confirm the observation of longitudinally polarized $W$s from the Higgs
boson decay.   

\section{Central Jet Veto}\label{sec:5}

In contrast to the $H\to  W^+W^-$ signal, the two central $W$s will be 
accompanied by close-by $b$-quark jets in the $t\bar tjj$ background
and, as we shall see, the $W+$~jets QCD background also produces more
observable central jets than the signal process. A veto on any 
{\it additional} central jet activity will thus substantially improve the 
signal to background ratio~\cite{VBHiggs,bpz,CMS,ATLAS}.   

In order to study the effects of semi-soft parton radiation for the 
Higgs signal and the background, we use the TSA of Section~\ref{sec:2d}, 
and thus we first need to estimate the TSA scales, $p_{TSA}$, for the 
various processes. For $W+4$~parton production we find that 
$p_{TSA}^{W+4~ jets} = 40.5~ {\rm GeV}$ reproduces the $W+3$~jets cross 
section of 8.36~pb in Table~\ref{Table_tag}. As discussed in 
Section~\ref{sec:2d} we use $p_{TSA}^{t\bar{t}}=42$~GeV for 
those $t\bar tjj$ events in which a $b$-quark arising in a top quark 
decay produces one of the forward tagging jets. For the Higgs signal, 
a separate estimate for the $m_{H}=800$~GeV and $m_{H}=100$~GeV cases 
gives two different values for $p_{TSA}$ which, if used, will lead to 
an incomplete subtraction of the continuum electroweak background. 
Instead, we match the difference between the two cross sections, 
$B\sigma(m_{H}=800~{\rm GeV})-B\sigma(m_{H}=100~{\rm GeV})$, to
the lower order cross section, which gives $p_{TSA}^{H+3jets}=7.9$~GeV.

The differences in $p_{TSA}$ values between the signal and the background 
reflect the different characteristics of the corresponding hard 
scattering processes. For the signal, the momentum transfer, $Q$, to the
color charges is given by the virtuality of the incident weak bosons 
in the longitudinal weak boson scattering process and, hence,
$Q_{signal}\approx p_T^{\rm tag}\alt M_{W}$. For the $W+4$~jets and 
the $t\bar{t}$ backgrounds, on the other hand, the corresponding scales 
are substantially larger, of the order of $E_T(W)$ or, even, the partonic 
center of mass energy. Very roughly, $p_{TSA}$, the jet transverse 
momentum scale at which multiple parton emission becomes important, 
is one to two orders of magnitude smaller than the momentum transfer 
of the corresponding hard scattering process.

\begin{table}[t]
\caption{ 
Signal and background cross sections $B\sigma$ in fb, before and after 
the veto of additional central jets.
% with $p_{T,j}^{soft}>20$ GeV, in the region 
%between the two tagging jets, at the LHC with $\protect\sqrt{s}=14$ TeV.
Cuts in the first column are the same as in the last column of 
Table~\protect\ref{Table_tag}, but cross sections are 
obtained within the TSA. The second 
column includes the dijet resolution and mass reconstruction efficiencies of
Table~\protect\ref{Table_Wjj} and columns three and four give cross sections
and expected event rates after the central jet veto of 
Eq.~(\protect\ref{veto_cut}). The Higgs signal cross section is defined as 
$B\sigma(m_{H})-B\sigma(m_{H}=100~{\rm GeV})$.   
}
\label{Table_veto}
\vskip 0.15in
\begin{tabular}{lcccc}
       & Hard cuts  & + jj resolution  & + central jet  & Number of events  \\ 
       & + soft jet & and $M_{jj}$ cut & veto & ${\cal L} = 100$~fb$^{-1}$  \\
         & (TSA)    &   efficiency     &         &  \\\hline
$m_{H}=800$ GeV  &  1.64   &  1.17   &  1.07   &  107   \\
$m_{H}=100$ GeV  &  0.17   &  0.10   &  0.08   &    8   \\
$m_{H}=1$   TeV  &         &         &  0.99   &   99   \\
$W$+4 jets       &  8.49   &  0.47   &  0.21   &   21   \\
$t\bar{t}$       &  1.55   &  1.10   &  0.12   &   12   \\[0.2in]
\underline{signal}:  &  &  & &\\
$m_{H}=800$ GeV &   1.47   &  1.07   &  0.99   &   99 \\
$m_{H}=1$ TeV   &          &         &  0.91   &   91 \\
\end{tabular}
\end{table}

The signal and the background cross sections which are obtained within the 
TSA are given in the first column of Table~\ref{Table_veto}. Within the 
Monte Carlo errors they agree with the $W+3$~jets, the $t\bar tjj$ and
the signal cross sections in the last column of Table~\ref{Table_tag}.
For all results presented in this section, the hadronically decaying $W$
is again assumed to decay into a single observable jet.
The resolution of this jet into two subjets and the effect of an 
invariant mass cut on this dijet system is then taken into account by 
multiplying with the efficiency factors given in Table~\ref{Table_Wjj}.
Like the Higgs signal, the $t\bar t$ background contains a 
predominantly longitudinally polarized $W$ which decays hadronically 
and the dijet resolution and dijet mass cut efficiencies for these decays
will be similar to the ones found for the signal. The signal 
efficiency of $0.71$ has therefore also been used for the $t\bar t$ 
background in the second column of Table~\ref{Table_veto}.
This procedure gives a conservative estimate of the top quark 
background since the combinatorial dilution of the $W\to jj$ peak,
due to $b$ quarks misidentified as $W$ decay jets, is not taken 
into account. 

\begin{figure}[t]
\vspace*{0.5in}            
\begin{picture}(0,0)(0,0)
\includegraphics{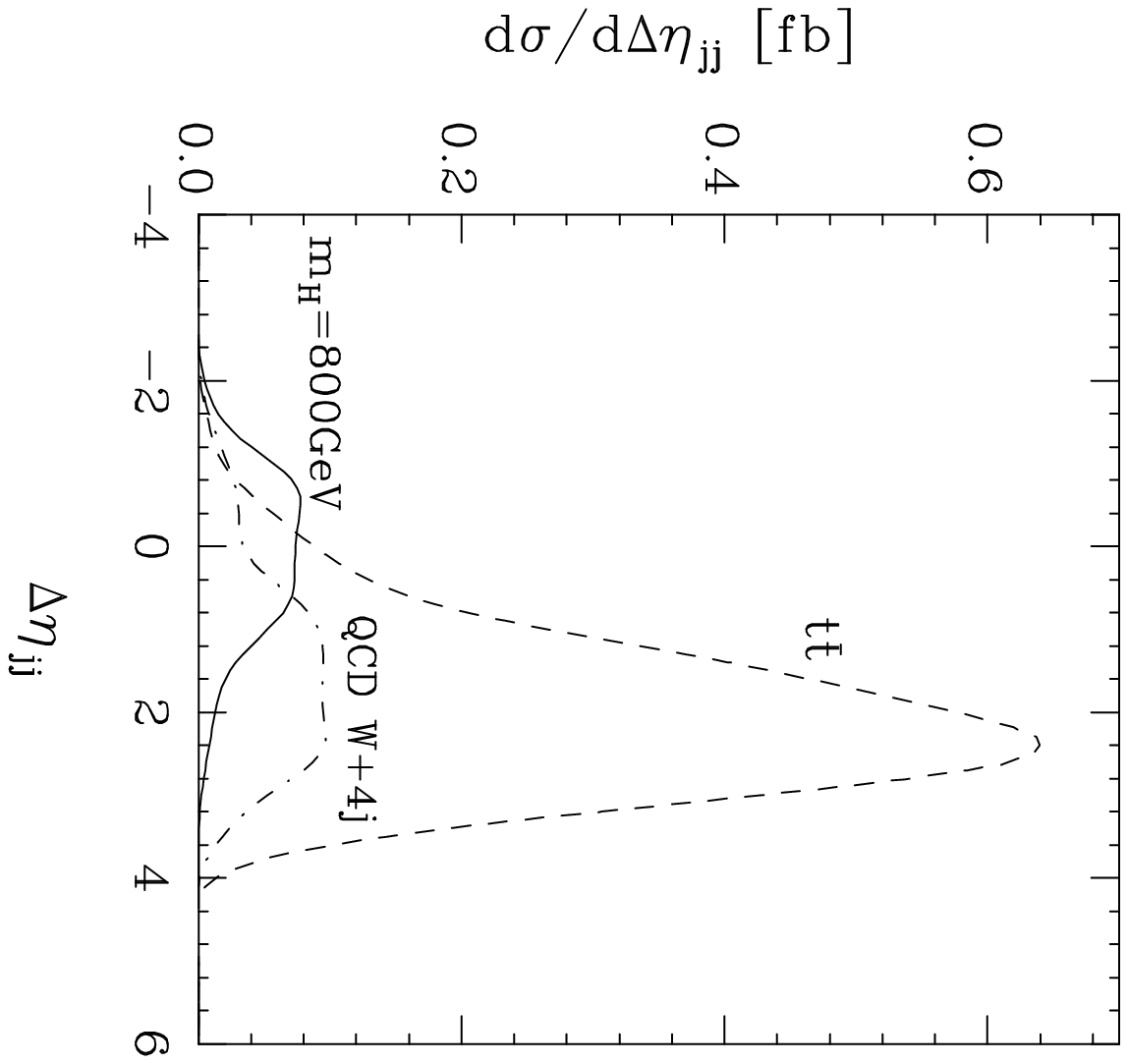}
\end{picture}
\vspace{9.3cm}
\caption{
Rapidity distance, $\Delta\eta_{jj}$, of secondary jets of $p_{Tj}>20$~GeV
from the closest forward tagging jet. Results are shown for the 
$m_H=800$~GeV ${\cal O}(\alpha_{s})$ electroweak processes (solid line), 
$t\bar tjj$ production (dash-dotted line) and QCD $W$+jets production 
(dotted line). Negative values of $\Delta\eta_{jj}$ correspond to 
secondary jets outside the rapidity interval formed by the two forward 
tagging jets.
\label{fig:yveto}
}
\vspace{5mm}
\end{figure}

For the Higgs signal the two forward tagging jets define the phase space
region in which to veto minijet activity. Color coherence favors
additional parton emission outside the rapidity range bounded by the two 
tagging jets. A good way to capture the differences between the signal and
the various backgrounds is by plotting the cross sections as a function of 
$\Delta \eta_{jj}$, the smallest relative distance, in units of 
pseudo-rapidity, between the extra jet and the two tagging jets,
\begin{equation}
\label{Delta_hjj}
      \Delta \eta_{jj} = {\rm sign} \left | \eta_{j}^{tag}({\rm closest}) -
                                            \eta_{j}^{soft} \right | \; .
\end{equation}
Here $\eta_{j}^{tag}({\rm closest})$ is the pseudo-rapidity of the 
forward tagging jet which is closest to the soft jet. The sign 
in Eq.~(\ref{Delta_hjj}) is chosen such that $\Delta \eta_{jj}$ is 
negative if the additional jet is outside the pseudo-rapidity interval
bounded by the two tagging jets  and positive otherwise.
The $\Delta \eta_{jj}$ distribution for additional jets with 
$p_{Tj}>20$~GeV is shown in Figure~\ref{fig:yveto}.
Both background distributions peak at $\Delta \eta_{jj}\sim 2$, indicating 
that the additional jet is predominantly emitted in the
central region, between the two forward tagging jets. This is in contrast
to the signal process, where an additional jet is 
emitted more forward than the tagging jets $\sim 50\%$ of the time.   
Because of the small scale governing gluon emission in the 
signal, only a small fraction of all signal events has an extra parton 
with transverse momentum in excess of 20~GeV. A good strategy is therefore 
to veto events with any additional jets between the two
tagging jets, i.e. events which have an additional jet satisfying
\begin{equation}
\begin{array}{ll}
\label{veto_cut}
 p_{Tj}^{soft}> p_{T,veto} = 20 {\rm GeV},~ & 
                \eta_{j}^{soft} \varepsilon \left [
                \eta_{j}^{tag1},\eta_{j}^{tag2} \right ]
\end{array}
\end{equation}
The rapidity requirement corresponds to a cut $\Delta \eta_{jj}>0$ in
Fig.~\ref{fig:yveto}.

\begin{figure}[t]
\vspace*{0.5in}            
\begin{picture}(0,0)(0,0)
\includegraphics{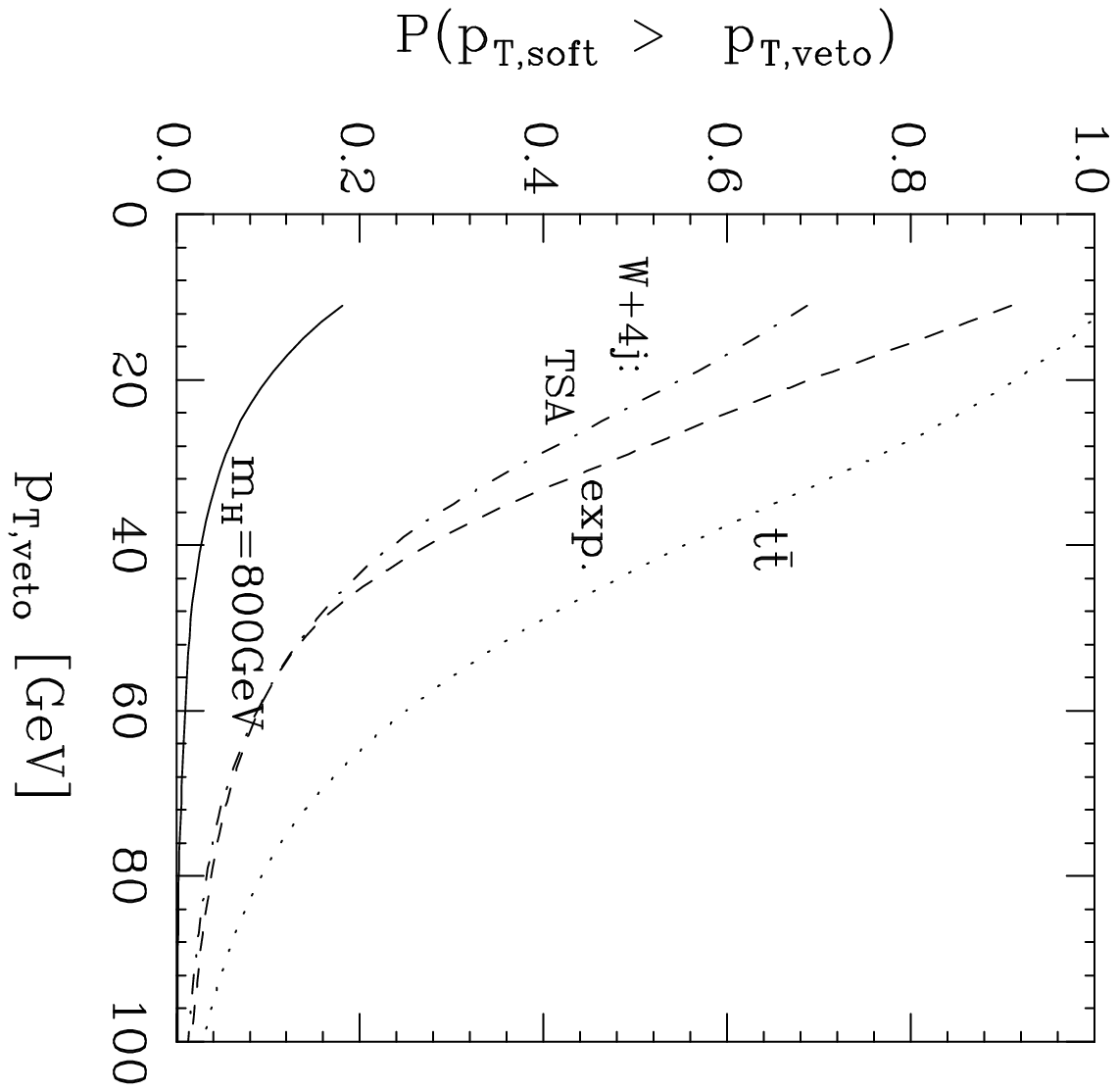}
\end{picture}
\vspace{9.3cm}
\caption{
Probability to find a veto jet candidate
above a transverse momentum $p_{T,\rm veto}$ between the two forward 
tagging jets. Results are derived in the TSA for the $m_H=800$~GeV electroweak
signal at ${\cal O}(\alpha_{s})$ (solid line), $t\bar tjj$ 
production (dotted line) and QCD $W$+jets production (dash-dotted line).
For QCD $W$+jets production the result for soft parton exponentiation is 
shown as the dashed line. See text for details.
\label{fig:vetoprob}
}
\vspace{5mm}
\end{figure}

The probability for finding a veto jet candidate depends strongly on the 
minimum transverse momentum, $p_{T,veto}$, of the additional jets.
Within the TSA this probability can be estimated by integrating  
$d\sigma_{TSA} / dp_{Tj}^{soft}$ over the allowed 
$p_{Tj}$ range. The result is then normalized to the lowest order 
cross section. Thus,
\begin{equation}
 P \left( p_{Tj}^{soft}> p_{T,veto} \right ) = {1 \over \sigma_{LO}}
   \int_{p_{T,veto}}^{\infty} 
   { d\sigma_{TSA} \over dp_{Tj}^{soft} } \cdot dp_{Tj}^{soft}
\end{equation}
with 
$\eta_{j}^{soft} \epsilon \left[ \eta_{j}^{tag1},\eta_{j}^{tag2}\right]$. 
This probability is shown in Figure~\ref{fig:vetoprob}.
At $p_{T,veto}=20$ GeV, the probability for finding a veto jet candidate 
in a signal event is below $10\%$, whereas there is substantial 
probability to find such a jet in background events ($\sim 55\%$ for the 
$W$+4~jets and $\sim 90\%$ for the $t\bar{t}jj$ backgrounds). 
For the $t\bar{t}$ background, the veto probability tends to 1 
as $p_{T,veto}\rightarrow 0$, due to the fact that
one of the two $b$-jets is almost always emitted in the veto region.
Within the TSA, the veto probability for the $W$+4 jets background remains
substantially less than 1, even if $p_{T,veto}\rightarrow 0$. 
This happens because in the 
TSA only one additional parton is emitted, with finite probability to be 
outside the veto region, as seen in Figure~\ref{fig:yveto}. Thus, at 
small $p_{T,veto}$ values, the TSA underestimates the veto probability.

An improved estimate of the veto probability at low $p_{T,veto}$ values
is obtained by assuming that in the soft region multiple parton emission 
is dominated by the emission of gluons, and that the gluon emission 
probability exponentiates. This model of multiple minijet emission predicts
a Poisson distribution for the multiplicity of additional minijets in 
hard scattering events. Indeed, recent CDF data are well described by
this ansatz~\cite{CDF,rsz}. Within this exponentiation model, the veto 
probability can then be estimated as~\cite{bpz}
\begin{equation}
\label{veto_expon}
 P_{exp} \left( p_{Tj}^{soft}> p_{T,veto} \right ) = 
   1 - \exp \left [ -
   {1 \over \sigma_{LO}} \int_{p_{T,veto}}^{\infty} 
   { d\sigma_{n+1} \over dp_{Tj}^{soft} } \cdot dp_{Tj}^{soft} \right ]
\end{equation}
where $ d\sigma_{n+1}/dp_{Tj}^{soft}$ is the unregularized $n+1$ parton
cross section, i.e. the higher order cross section without the 
truncated shower approximation. 

A veto probability estimate based on Eq.~(\ref{veto_expon}) is also 
shown in Figure~\ref{fig:vetoprob}, for the $W$+4 jets background.
At $p_{T,veto}=20$ GeV, the TSA underestimates the veto probability by 
$20\%$. For large values of $p_{T,veto}$ ($p_{T,veto}>50$ GeV), the two 
calculations give essentially the same veto probability. 
In the following estimates for the observability of a heavy Higgs signal at
the LHC, the more conservative TSA results
for the veto probability are used.

Signal and the background cross sections, after applying the central jet 
veto with $p_{Tj}>p_{T,veto}=20~{\rm GeV}$, are given in the third column
of Table~\ref{Table_veto}.
As expected, the jet veto is extremely effective in removing 
the $t\bar{t}$ background, due to the presence of the two $b$-jets. 
The integrated cross section is reduced by one order of magnitude 
and is now only $\sim 12\%$ of the signal cross section.
The $W$+4 jets background is reduced by a factor 2.2. These background 
reductions are achieved with a very high efficiency for retaining the signal,
with approximately $91\%$ of the signal events passing the veto criterion.

After the veto of additional jets in the central region, the signal
cross section rate is a factor of 2.5 larger than the combined background 
rate. Assuming an integrated luminosity of ${\cal L}=100~{\rm fb}^{-1}$,
the expected number of events for the Higgs signal and for the background 
are given in the last column of Table~\ref{Table_veto}.
At $m_{H}=800$ GeV, 99 signal events are expected with a total background 
of 41 events.
These numbers indicate that the Higgs boson can be discovered in the 
$H\rightarrow WW\rightarrow l\nu jj$ decay mode, for Higgs masses up 
to $m_{H}=1$ TeV.

\section{Discussion and Conclusions}\label{sec:6}

Our analysis of the $H\to WW\to l\nu jj$ decay mode of a heavy Higgs boson 
is based on complete tree level QCD calculations of the cross sections 
for the $qq\to qqWW$ signal as well as for the $W+3,4$~jets 
and $t\bar tjj$ background processes. Full QCD matrix elements provide the
most reliable predictions for event features like hard jet distributions, 
the momentum scales governing the emission probability of additional 
soft jets, and the angular distributions of such additional jets. With 
QCD matrix elements for $(W\to l\nu)(W\to jj)+2,3$~jet events for the signal,
$W+3,4$~jet events for the QCD background 
and $(t\to Wb)(\bar t\to W\bar b)+2$~jet production for the top-quark 
background we have analyzed optimal criteria for double forward jet tagging, 
the expected resolution and expected background suppression when searching 
for a $W\to jj$ invariant mass peak, prospects for measuring the longitudinal 
polarization of the hadronically decaying $W$ of the Higgs boson signal, and 
we have studied momentum scales and angular distributions of additional 
soft jet emission which would be affected by a central jet veto.

Previous analyses by the ATLAS~\cite{ATLAS} and CMS~\cite{CMS} Collaborations
used parton shower programs like PYTHIA~\cite{pythia} instead, which 
give a more detailed description of other aspects of signal and background
events, like particle content, higher soft jet multiplicities, and the 
presence of an underlying event. Carrying the simulation to the particle level
also allows for a more realistic assessment of detector response.
With fairly similar acceptance cuts on the Higgs
decay products and using double forward jet tagging and central jet vetoing
techniques these studies arrived at qualitatively the same answer: that a 
heavy Higgs boson can be discovered in the $H\to WW\to l\nu jj$ decay mode.
It is reassuring that also quantitatively the agreement is excellent. The 
predicted  rates for the Higgs boson signal and the various backgrounds 
found in the ATLAS and CMS analyses are somewhat smaller than
ours, by up to a factor two. While this general trend is expected from 
including detector efficiencies, our central-jet-vetoing and 
forward-jet-tagging criteria differ sufficiently from those used in 
Refs.~\cite{CMS,ATLAS} that this agreement may be fortuitous to some extent. 
Since we do not have the parton level cross sections for these analyses, a more
explicit and quantitative comparison with the ATLAS and CMS studies is not
feasible. A few differences are noteworthy, however.

Both the CMS and the ATLAS analysis
find suppression of the $W+$~jets background by a factor $\approx 3$ from
a central jet veto, but with very different values of $p_{T,veto}=40$~GeV 
(CMS) and $p_{T,veto}=15$~GeV (ATLAS), and ATLAS reports a decrease to a factor
2.5 when $p_{T,veto}=40$~GeV is used, at high luminosity. Our analysis 
indicates that the veto probability should vary more strongly with 
$p_{T,veto}$ (see Fig.~\ref{fig:vetoprob})
and this question deserves further study. 
Another difference is the transverse energy requirement for the two tagging
jets: our choice of transverse energy threshold for the tagging jets,
$E_{Tj}(tag)>50$~(30)~GeV, is 
motivated by the signal distributions in Fig.~\ref{fsignalpt}. 
It is considerably harder than the values $E_{Tj}(tag)>15$~GeV (ATLAS) and
$E_{Tj}(tag)>10$~GeV (CMS) used in the Technical Proposals of the two 
detectors~\cite{CMS,ATLAS}. Thus our choice definitely is conservative.

Neither of the presently available analyses will be definitive by the time
the LHC experiments start taking data. The tools presented here,
based on state-of-the-art QCD matrix elements, can be used to calibrate
parton shower Monte Carlo programs. Combined with full
detector simulations they will provide the more reliable predictions for
longitudinal weak boson scattering signals and background processes
which are needed to understand LHC data.

%\newpage
\acknowledgements
We are grateful to A.~Stange for making his $t\bar t jj$ Monte Carlo program 
available to us and thank E.~Mirkes for many useful conversations.
This research was supported by the University of Wisconsin 
Research Committee with funds granted by the Wisconsin Alumni Research 
Foundation and by the U.~S.~Department of Energy under Grant 
No.~DE-FG02-95ER40896.

\newpage             %%%%%%%%  REFERENCES %%%%%%%%%%%

\end{document}